\documentclass[aps,prl,floatfix,twocolumn,superscriptaddress]{revtex4-1}
\usepackage{amsmath,amssymb}
\usepackage{graphicx}
\usepackage{epsfig}
\usepackage{psfrag}
\usepackage[usenames]{color}
\usepackage{xcolor}
\usepackage{hyperref}
\usepackage{soul,color}
\usepackage{physics}
\hypersetup{colorlinks=true,citecolor={blue},linkcolor={blue},urlcolor={blue}}

\begin{document}

\title{Bogolon--mediated electron scattering in graphene \\in hybrid Bose--Fermi systems}


\author{Meng~Sun}
\affiliation{Center for Theoretical Physics of Complex Systems, Institute for Basic Science (IBS), Daejeon 34126, Korea}
\affiliation{Basic Science Program, Korea University of Science and Technology (UST), Daejeon 34113, Korea}

\author{K.~H.~A.~Villegas}
\affiliation{Center for Theoretical Physics of Complex Systems, Institute for Basic Science (IBS), Daejeon 34126, Korea}

\author{V.~M.~Kovalev}
\affiliation{A.~V.~Rzhanov Institute of Semiconductor Physics, Siberian Branch of Russian Academy of Sciences, Novosibirsk 630090, Russia}
\affiliation{Department of Applied and Theoretical Physics, Novosibirsk State Technical University, Novosibirsk 630073, Russia}

\author{I.~G.~Savenko}
\affiliation{Center for Theoretical Physics of Complex Systems, Institute for Basic Science (IBS), Daejeon 34126, Korea}
\affiliation{Basic Science Program, Korea University of Science and Technology (UST), Daejeon 34113, Korea}

\date{\today}

\begin{abstract}
We report on an unconventional mechanism of electron scattering in graphene in hybrid Bose-Fermi systems.
We study energy--dependent electron relaxation time, accounting for the processes of emission and absorption of a Bogoliubov excitation (a bogolon).
Then using the Bloch-Gr\"uneisen approach, we find the finite-temperature resistivity of graphene and show that its principal behavior is $\sim T^4$ in the limit of low temperatures and linear at high temperatures. 
We show that bogolon-mediated scattering can surpass the acoustic phonon--assisted relaxation. 
It can be controlled by the distance between the layers and the condensate density, giving us additional degrees of freedom and a useful tool to render electron mobility by the sample design and external pump.
\end{abstract}	

\maketitle


\textit{Introduction.---}Electron scattering in solid-state nanostructures plays crucial role in their two-dimensional transport~\cite{Kawamura:1992aa,Hwang:2008aa}, dramatically modifying electric conductivity. 
Conventionally there exist two principal mechanisms of electron scattering: disorder or impurity--mediated~\cite{Jena:2007aa,Gibbons:2009aa} and lattice phonon--mediated~\cite{Kawamura:1992aa} scatterings. The former processes are more pronounced at low ambient temperatures.
In the case of an attracting impurity, electrons can be captured, thus the number of electrons decreases, while repulsive centers make the electron mean free path and scattering time decrease~\cite{Shi:2012aa,Bourgoin:1992aa,Eshchenko:2002aa,Palma:1995aa,Boev:2018ab}. 
With the increase of temperature, electron scattering accompanied by the emission and absorption of acoustic and optical phonons of the crystal lattice becomes more efficient~\cite{Gummel:1955aa, Lax:1960aa, Abakumov:1976aa} and at some point dominant.

Conventional scattering mechanisms also take place in hybrid structures of various new kinds, which are in focus of modern research~\cite{Cotleifmmode-telse-tfi:2016aa,Laussy:2010aa}. Hybrid systems consist of two-dimensional spatially separated layers, containing electrons in a two-dimensional electron gas (2DEG) phase and bosons, such as direct and indirect (dipolar) excitons, exciton polaritons, or the Cooper pairs in superconductors~\cite{Villegas:2018aa}. 
In these systems, the research is, on the one hand, devoted to high-temperature boson-mediated  superconductivity~\cite{Skopelitis:2018aa} and other condensation phenomena in interacting structures, including the Mott phase transition from an ordered state to electron-hole plasma~\cite{Kochereshko:2016aa}. On the other hand, in such systems there can appear new mechanisms of scattering of fermions in the 2DEG, thus modifying the temperature dependence of the kinetic coefficients. These arguments explain the motivations to study electron transport in hybrid systems.

In this Letter, we show that in hybrid Bose-Fermi systems, which consist of spatially separated 2DEG in graphene layer and an exciton gas, interacting via the Coulomb forces~\cite{Boev:2016aa,Kochereshko:2016aa, Matuszewski:2012aa}, there appears a counterpart to the phonon-mediated scattering, when the gas of bosons is condensed~\cite{Kovalev:2011aa, Kovalev:2013aa, Batyev:2014aa}. Two-dimensional condensation has been reported in various solid state systems~\cite{Butov:2003aa,Kasprzak:2006aa,Schneider:2013aa}. 
There the lattice vibrations turn out to be not the only \textit{sound} available. In the presence of a Bose-Einstein condensate (BEC)~\cite{Butov:2003aa}, there come into play other excitations, commonly referred to as Bogoliubov quasiparticles or bogolons which have linear dispersion at small momenta.

We ascertain, that an additional principal mechanism of electron scattering appears, stemming from the interlayer electron-exciton interaction: bogolon-mediated scattering. And the difference between acoustic phonons-related and bogolon-assisted scattering is more than just the magnitude of the sound velocity. 
%
The dependence of the bogolon--mediated resistivity of graphene on temperature is $\sim T^4$ at low temperatures and $\sim T$ in the high-temperature limit. In contrasts, a precise calculation of the acoustic phonon--mediated resistivity in graphene shows $\rho\propto T^\alpha$ at low temperatures with $\alpha\sim6$~\cite{Kaasbjerg:2012aa}.
Moreover, the phonon-mediated scattering is vulnerable to the screening effects~\cite{Hwang2007PRB75205418}.
This makes a great deal of difference between the two mechanisms of scattering, and one can surpass the other.

\begin{figure}[!t]
\includegraphics[width=0.49\textwidth]{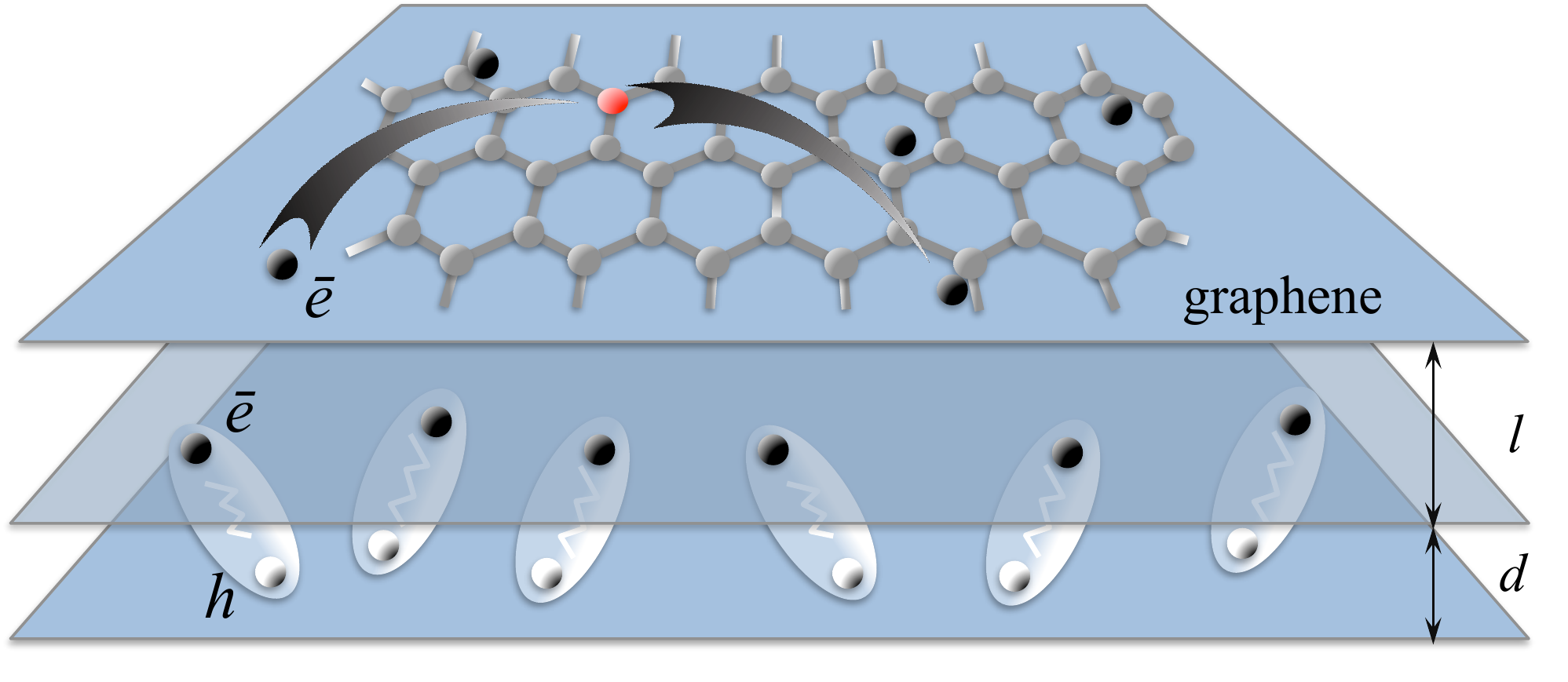}
\caption{System schematic. Graphene, located at the distance $l$ from a two-dimensional dipolar exciton gas, residing in two parallel layers, which are at the distance $d$ from each other. Particles couple via the Coulomb interaction.}
\label{fig:1}
\end{figure}

\textit{System schematic.---}Let us consider a hybrid system consisting of a graphene layer, separated by the distance $l$ from a double quantum well, containing a dipolar exciton gas, where the distance between the layers of electrons and holes is $d$ (see Fig.~\ref{fig:1}).
The electron-exciton interaction can be described by the Hamiltonian,
\begin{equation}\label{eq.1}
V=\int d\mathbf{r}\int d\mathbf{R}\Psi^\dag_\mathbf{r}\Psi_\mathbf{r}g\left(\mathbf{r}-\mathbf{R}\right)\Phi^\dag_\mathbf{R}\Phi_\mathbf{R},
\end{equation}
where $\Psi_\mathbf{r}$ and $\Phi_\mathbf{R}$ are the quantum field operators of electrons and excitons, correspondingly, $g\left(\mathbf{r}-\mathbf{R}\right)$ is the Coulomb interaction between an electron and an exciton, $\mathbf{r}$ is electron coordinate within the graphene plane, and $\mathbf{R}$ is the exciton center-of-mass coordinate. 
We will disregard the internal structure of excitons and only concentrate on their collective motion.

\begin{figure}[htb]
    \centering
    \includegraphics[width=0.49\textwidth]{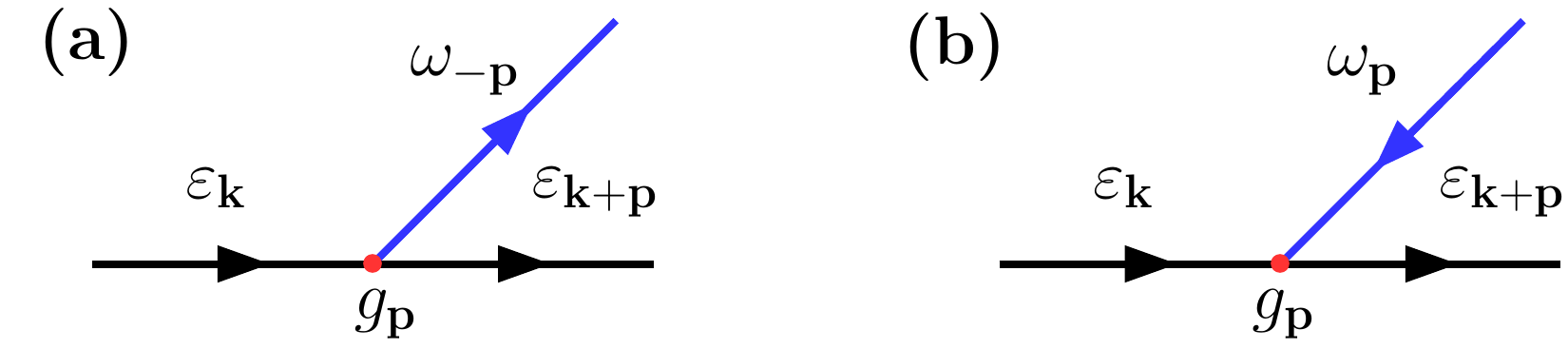}
    \caption{Schematic of the electron scattering, mediated by the bogolon emission (a) and absorption (b) processes.}
    \label{fig:2}
\end{figure}

If the excitons condense, we can use the model of weakly non-ideal interacting Bose gas. Then the exciton field operators read $\Phi_\mathbf{R}=\sqrt{n_c}+\phi_\mathbf{R}$, where  $n_c$ is the  condensate density, thus we separate the condensed and non-condensed particles.
Substituting this in Eq.~\eqref{eq.1} and taking into account the selection rules, we find the electron--bogolon interaction potential,
\begin{equation}\label{eq.2}
V = \sqrt{n_c}\int d\mathbf{r}\Psi^\dag_\mathbf{r}\Psi_\mathbf{r} \int d\mathbf{R}g\left(\mathbf{r}-\mathbf{R}\right)\left[\varphi^\dag_\mathbf{R}+\varphi_\mathbf{R}\right].
\end{equation}
Furthermore we take the Fourier transform of the operators in~\eqref{eq.2}, using
\begin{equation}\label{eq.3}
\varphi^\dag_\mathbf{R}+\varphi_\mathbf{R}=\sum_{\mathbf{p}} e^{i\mathbf{pR}} \left[(u_\mathbf{p}+v_{-\mathbf{p}})b_\mathbf{p}+(v_\mathbf{p}+u_{-\mathbf{p}})b^\dag_{-\mathbf{p}}\right], 
\end{equation}
where $b^\dag_{\mathbf{p}}$ and $b_{\mathbf{p}}$ are the creation and annihilation operators of the bogolons, and the coefficients read~\cite{Giorgini:1998aa}
\begin{eqnarray}\label{eq.4}
u^2_{\mathbf{p}}&=&1+v^2_{\mathbf{p}}=\frac{1}{2}\left(1+\left[1+\frac{(Ms^2)^2}{\omega^2_{\mathbf{p}}}\right]^{1/2}\right),\\\nonumber
&&u_{\mathbf{p}}v_{\mathbf{p}}=-\frac{Ms^2}{2\omega_{\mathbf{p}}}.
\end{eqnarray}
Here $M$ is the exciton mass, $s=\sqrt{\kappa n_c/M}$ is the sound velocity of bogolons,
$\omega_k=sk(1+k^2\xi^2)^{1/2}$ is their spectrum, $\kappa=e_0^2d/\epsilon$ is the Fourier image of the exciton--exciton interaction strength, $e_0$ is electron charge, $\epsilon$ is the dielectric function and $\xi=\hbar/(2Ms)$ is the healing length. 
Combining Eqs.~\eqref{eq.2} and~\eqref{eq.3} yields
\begin{equation}\label{eq.5}
V = \sqrt{n_c} \sum_{\mathbf{k,p}} g_\mathbf{p} \left[ \left( v_\mathbf{p} + u_\mathbf{-p} \right)b^\dagger_\mathbf{-p} + \left( u_\mathbf{p} + v_\mathbf{-p}\right)b_\mathbf{p} \right] c^\dagger_\mathbf{k+p} c_\mathbf{k}, 
\end{equation}
where $g_\mathbf{k}= e^2_0 d e^{-kl}/(2\epsilon)$ is the Fourier image of the electron-exciton interaction.
The schematic of the processes~\eqref{eq.5} is presented in Fig.~\ref{fig:2}, showing scattering of an electron, mediated by an absorption or emission of a bogolon.
%
%
%


\textit{Transport of particles.---}Further we use the Boltzmann transport theory \cite{Kawamura:1990aa} to calculate the resistivity of electrons in graphene, which is given by
%
\begin{equation}\label{eq.6}
\rho^{-1} = e_0^2 D\left(E_F\right)\frac{v_F^2}{2}\langle \tau \rangle,
\end{equation}
where $v_F$ is the Fermi velocity, $E_F$ is the Fermi energy, and the density of states of graphene at the Fermi level reads $D\left(E_F\right) =\left(g_s g_v/2\pi \hbar^2\right)E_Fv_F^2$, where $g_{s,v}=2$ are the spin and valley g-factors, respectively. 
We can write the energy-averaged relaxation time as
\begin{equation}\label{eq.7}
\langle \tau \rangle = \frac{\int d \varepsilon D\left(\varepsilon \right) \tau\left( \varepsilon \right) \left[ -\frac{df^0\left(\varepsilon\right)}{d \varepsilon}\right]}{\int d \varepsilon D\left(\varepsilon\right) \left[ -\frac{df^0\left(\varepsilon\right)}{d \varepsilon} \right]},
\end{equation}
where $f^0\left(\varepsilon\right)=\{\exp[(\varepsilon-\mu)/(k_BT)]\}^{-1}$ is the Fermi distribution function, $\mu$ is the chemical potential, $k_B$ is the Boltzmann constant, and the energy--dependent inverse relaxation time reads
\begin{equation}\label{eq.8}
   \frac{1}{\tau\left(\varepsilon\right)} = \sum_\mathbf{k'}\left(1 -\cos\theta_\mathbf{kk'}\right) W_\mathbf{kk'} \frac{1-f^0\left(\varepsilon'\right)}{1-f^0\left(\varepsilon\right)},
\end{equation}
where $\theta_\mathbf{kk'}$ is the scattering angle between $\mathbf{k}$ and $\mathbf{k'}$, $\varepsilon=\hbar v_F\abs{\mathbf{k}}$ is the dispersion of graphene and $W_\mathbf{kk'}$ is the probability of transition from an initial electron state $\mathbf{k}$ to the final state $\mathbf{k}'$, given by 
\begin{equation}\label{eq.9}
    W_\mathbf{kk'}=\frac{2\pi}{\hbar}\sum_\mathbf{q} \abs{C_\mathbf{q}}^2 \Delta\left(\varepsilon, \varepsilon'\right).
\end{equation}
Here $C_\mathbf{q}$ is the scattering matrix element, and 
\begin{equation}\label{eq.10}
   \Delta\left(\varepsilon,\varepsilon'\right)=N_q\delta\left(\varepsilon-\varepsilon'+\hbar\omega_q\right)+\left(N_q+1\right)\delta\left(\varepsilon-\varepsilon'-\hbar\omega_q\right) ,
\end{equation}
where $N_q=\{\exp[\hbar\omega_q/(k_BT)]-1\}^{-1}$ is the Bose distribution function.
Summing up, the energy--dependent relaxation time reads
\begin{eqnarray}\label{eq.11}
\frac{1}{\tau\left(\varepsilon\right)} &=& \frac{e_0^4 d^2 n_c}{8 \pi\epsilon^2 \hbar} \int d \mathbf{k'}\left(1-\cos_{\mathbf{kk'}}\right) \int d \mathbf{q} e^{-2\abs{q}l}\abs{u_\mathbf{q}+v_\mathbf{q}}^2 \nonumber \\
&\times& \frac{1-f^0\left(\varepsilon'\right)}{1-f^0\left(\varepsilon\right)}\Delta\left(\varepsilon,\varepsilon'\right)\delta \left(\mathbf{q}-\mathbf{k}+\mathbf{k'}\right).
\end{eqnarray}
Using Eq.~\eqref{eq.4} and assuming linear dispersion of bogolons $\omega_\mathbf{q}=s\abs{\mathbf{q}}$ (which is legitimate at $q\ll \xi^{-1}$ ), we find:
\begin{eqnarray} \label{eq.12}
\frac{1}{\tau\left(\varepsilon\right)} &=& \sum_{n=1,2}\frac{e_0^4 d^2 n_c }{8\pi \epsilon^2 \hbar^3 v_F^2} \int_0^{2\pi} d\theta \varepsilon_n \left( 1-cos\theta\right)  \frac{1-f^0\left(\varepsilon_n \right) } {1-f^0\left(\varepsilon\right)} \nonumber \\ 
&\times& e^{-2l\lambda}\left( \sqrt{1+\frac{M^2s^2}{\hbar^2\lambda^2}} - \frac{Ms}{\hbar \lambda}\right) \frac{N_\lambda + \delta_{n,2}}{|F'_n \left( \varepsilon_n \right) |} ,
\end{eqnarray}
where $\lambda\equiv\abs{\mathbf{k}-\mathbf{k'}} = \sqrt{k^2+k'^2-2kk'\cos\theta}$, thus it is a function of $k$, $k'$ and $\theta$; $\varepsilon_n$ are two roots of the equation $F_{1,2} \left(\varepsilon'\right)=\varepsilon-\varepsilon'\pm \hbar \omega_\lambda=0$, $F'_{n}\left(\varepsilon'\right)$ is its first derivative, and $\delta_{n,2}$ is the Kronecker delta. Specifically, $n=2$ corresponds to the bogolon emission process. 
Substituting~\eqref{eq.12} in the average lifetime~\eqref{eq.7}, we can numerically calculate the conductivity~\eqref{eq.6}. However, let us first analytically consider the limiting cases of high and low temperatures.


\textit{High-temperature limit.---}Let us analise Eq.~\eqref{eq.12} and find the principal dependence of conductivity on $T$ at high temperatures, $T_{BG} \ll T \ll E_F/k_B$, where we denote the Bloch-Gr\"uneisen temperature as $T_{BG}=2\hbar sk_F/k_B$.
Since $T \gg T_{BG}$, we have $\hbar \omega_{\mathbf{q}} \ll k_B T$. 
In this case, the Bose-Einstein distribution can be approximated as $N_q \sim k_B T / \hbar \omega_\mathbf{q}$, and $\Delta\left(\varepsilon,\varepsilon'\right) = \left( 2k_B T / k_B T \right) \delta\left( \varepsilon -\varepsilon' \right)$.
Then we find the energy--dependent relaxation time,
\begin{eqnarray} \label{eq.13}
\frac{1}{\tau\left(\varepsilon\right)} &=& \frac{e_0^4d^2k_BT}{8\pi^2\epsilon^2\hbar^2v_F^2} \int^{2\pi}_0 d\theta \left( 1- \cos\theta\right) \varepsilon  \nonumber\\
&\times& e^{-\lambda l}\left( \sqrt{ \frac{1}{s^2 \lambda^2} + \frac{M^2}{\hbar^2 \lambda^4}} -\frac{M}{\hbar \lambda} \right) .
\end{eqnarray}
One should notice that the integral in Eq.~\eqref{eq.13} is temperature-independent.
Under the limit $T\ll E_F/k_B$, the contribution from the Fermi energy in Eq.~\eqref{eq.7} is dominant.
This gives us $\langle \tau \rangle \approx \tau\left(E_F\right)\sim T^{-1}$. Substituting this expression in Eq.~\eqref{eq.6}, we find that the resistivity depends on the temperature linearly, as in the case of phonon-assisted relaxation~\cite{Hwang:2008aa}.
Indeed the temperature should still be smaller than the exciton condensation temperature. Otherwise, bogolon-mediated relaxation does not exist.


\textit{Low-temperature limit.---}
To investigate the principal $T$-dependence of resistivity at low temperatures, we will use the Bloch-Gr\"uneisen formalism, described in~\cite{Ziman:2001aa, Zaitsev:2014aa}.
We start from the Boltzmann equation
\begin{equation}\label{1}
e_0\textbf{E}\cdot\frac{\partial f}{\hbar\partial \textbf{p}}=I\{f\},
\end{equation}
where $f$ is the electron distribution, $\mathbf{p}$ is the wave vector ($p\equiv\abs{\mathbf{p}}$), $\mathbf{E}$ is the perturbing electric field, and $I\{f\}$ is the collision integral (see Supplemental Material~\cite{[{See Supplemental Material at [URL] for the details of derivation of the Bloch-Gr\"uneisen formula}]SMBG} for the explicit form of $I$ and other details of derivation).
For relatively weak electric fields, $f$ can be expanded as
\begin{eqnarray}
f=f^0(\varepsilon_p)-\left(-\frac{\partial f^0}{\partial\varepsilon_p}\right)f^{(1)}_\textbf{p},
\end{eqnarray}
where the correction $f^{(1)}_\textbf{p}$ has the dimensionality of energy.
Without loss of generality, we put the electric field to direct along the $x$-axis 
and use the ansatz 
\begin{eqnarray}
f^{(1)}_\textbf{p}= v_F\frac{e_0E_xp_x}{k_F}\tau(\varepsilon_p).
\end{eqnarray}
%
After some algebra, we find the resistivity in the form
\begin{eqnarray}
\label{rhoV1}
\rho\propto\frac{1}{\tau_0}&=&\frac{\hbar\xi_I^2}{8\pi^2k_FM}\frac{1}{k_BT}\int_0^\infty dq q^4e^{-2ql}q(\Gamma_--\Gamma_+)_{k_F}\nonumber\\
&{}&{}\times N_q(1+N_q),
\end{eqnarray}
where $\tau_0$ is an effective scattering time, $\xi_I= e_0^2d\sqrt{n_c}/2\epsilon$, 
\begin{eqnarray}
\label{gamma2}
\Gamma_\pm=\frac{2|v_Fk_F\pm sq|(2v_Fsk_F-v_F^2q)}{\hbar v_F^3k_F q\sqrt{\pm 4k_Fsv_Fq+4k_F^2v_F^2-v_F^2q^2}},
\end{eqnarray}
and the subscript $k_F$ in the expression $(\Gamma_--\Gamma_+)_{k_F}$ in~\eqref{rhoV1} means that all the electron wave vectors $p$ are to be substituted by $k_F$ there.


%
%
%
\begin{figure}[!t]
    \centering
    \includegraphics[width=0.47\textwidth]{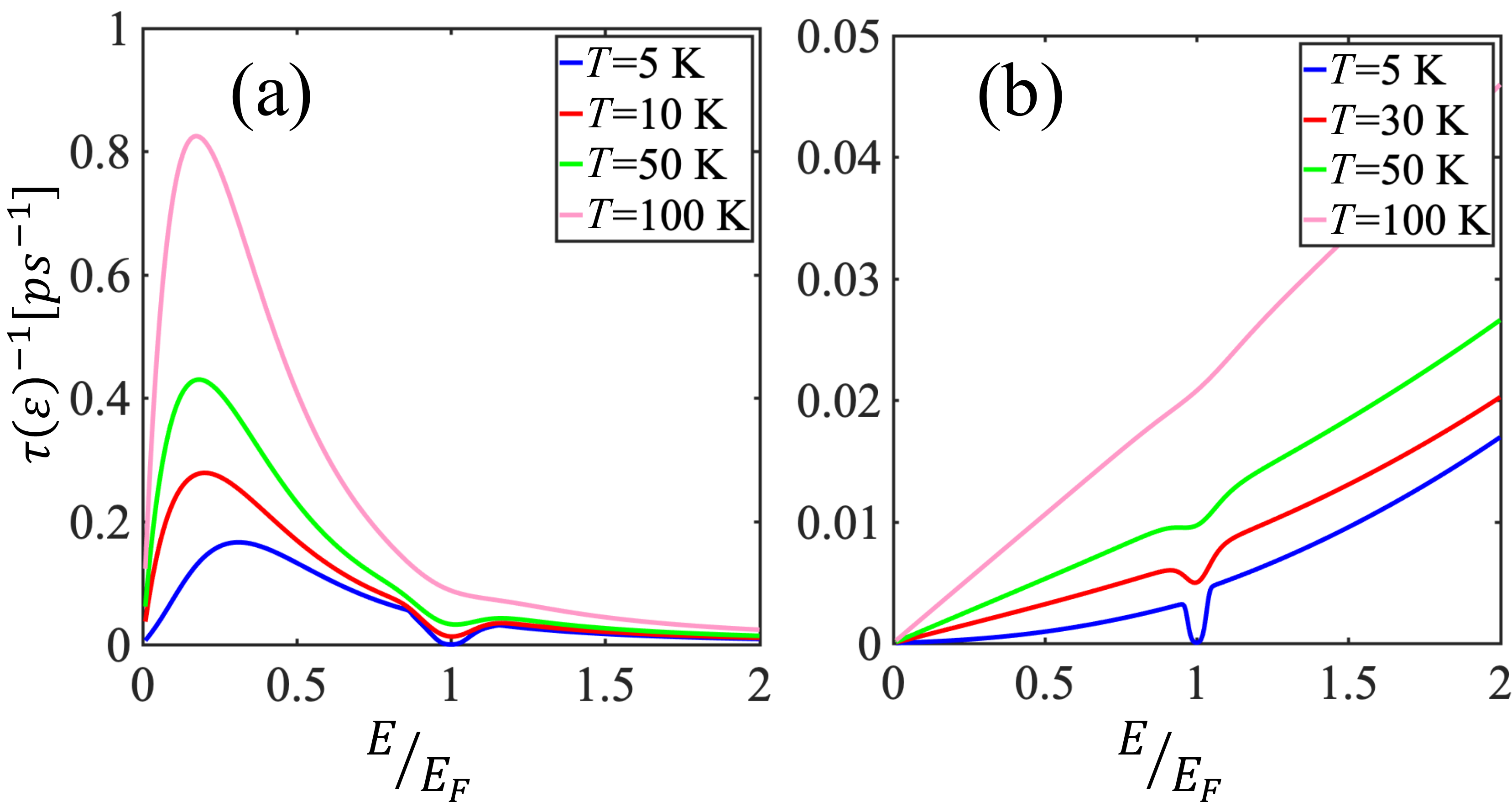}
    \caption{Energy--dependent inverse relaxation time of electrons for the single-bogolon (emission and absorption) processes (a) and phonon-assisted processes (b) for different temperatures;  $n_c=10^{11}$ cm$^{-2}$ and thus $s\approx 7\times 10^6$~cm/s.}
    \label{fig:3}
\end{figure}
%
%
%
%
%
%
\begin{figure}[!t]
    \includegraphics[width=0.49\textwidth]{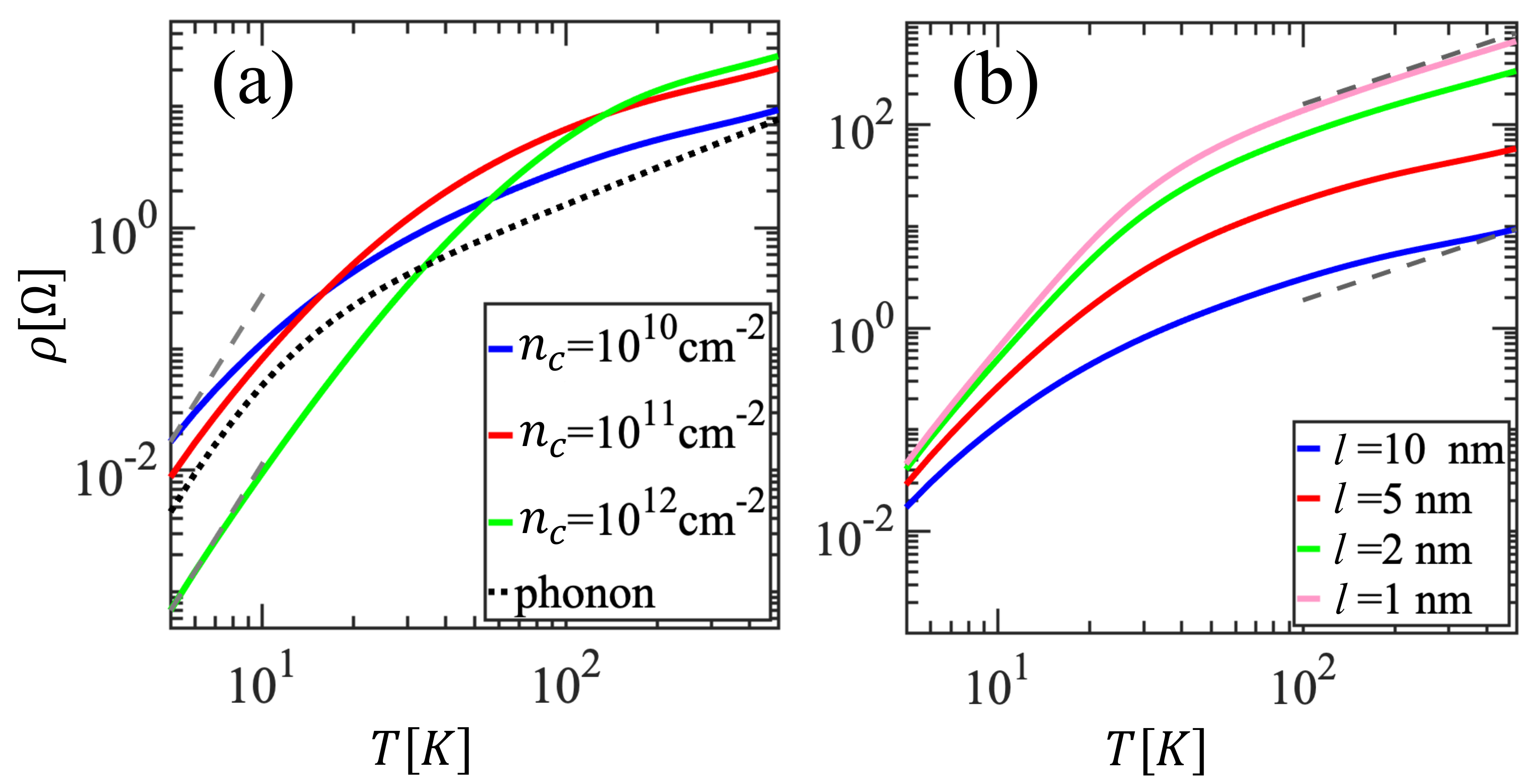}
    \caption{Bogolon-mediated resistivity of graphene as a function of temperature in the range $5$--$500$~K for different densities of particles in the condensate $n_c$, $l=10$ nm (a) and for different interlayer distances $l$, $n_c=10^{10}$~cm$^{-2}$ (b). 
    The dashed gray lines stand for the low- and high-temperature analytics, indicating the $\sim T^4$ and $\sim T$  behavior, respectively.  
    The black dotted line in panel (a) shows the phonon-mediated resistivity for comparison. }
    \label{fig:4}
\end{figure}

For temperatures much lower than the Bloch-Gr\"{u}neisen temperature, we find the following expression,
\begin{eqnarray}
\frac{1}{\tau_0}=\frac{I_0\xi_I^2k_F^2}{4\pi^2\hbar \alpha^4v_F^2M}\left(\frac{k_BT}{E_F}\right)^4,
\end{eqnarray}
where $I_0\approx 26.2$ is a dimensionless factor.
%
In terms of the resistivity,
\begin{eqnarray}
\label{EqResist}
\rho=\frac{\pi\hbar^2}{e_0^2E_F}\frac{1}{\tau_0}=(1.0\times 10^6\;\Omega)\left(\frac{k_BT}{E_F}\right)^4.
\end{eqnarray}
%
%
In this esteem, we used a dimensionless parameter $\Tilde{l}=lk_BT/(\hbar s)$ which is determined by the interlayer distance $l$, the sound velocity $s$ (which is in turn determined by the condensate density), and temperature; and we used the condition $\Tilde{l}\ll 1$ to get an analytical dependence at low $T$.

For temperatures far less than the room temperature ($k_BT_R\approx 26$ meV) we  have $\Tilde{l}\ll 1$. 
If $T\ll T_{BG}$, where $T_{BG}\ll E_F/k_B$ since $s\ll v_F$, we find the precise form of what we mean by \textit{low temperatures}: $k_BT/E_F<10^{-2}$. For typical $E_F\sim 10^{-1}$ eV, this gives $T_{BG}=183$ K and  $T<18$ K (for the particular range of distances between the layers $l$ up to $50$ nm).

Onwards, it is interesting and instructive to compare the formula~\eqref{EqResist}, rewritten in a different form,
\begin{eqnarray}
\frac{1}{\tau_0}=\frac{5I_0e_0^6}{8\pi^2\epsilon^2v_F^2}\frac{n_cd}{M}\frac{1}{E_Fk_F}\left(\frac{k_BT}{\hbar s}\right)^4,
\end{eqnarray}
with the phonon-mediated scattering case~\cite{Hwang:2008aa}
\begin{eqnarray}
\label{phonon}
\frac{1}{\tilde{\tau}_0}=\frac{D^24!\zeta(4)}{2\pi\rho_mv_{ph}}\frac{1}{E_Fk_F}\left(\frac{k_BT}{\hbar v_{ph}}\right)^4,
\end{eqnarray}
where $\rho_m$ is the density of graphene, $\zeta$ is the Riemann zeta-function. We see that both the inverse times have the same $T$-dependence at low temperatures with the phonon velocity $v_{ph}$ replaced by the sound speed $s$ in the bogolon-mediated scattering case.
%
%
%
%
%

This is not the end of the story yet. The result presented in~\cite{Hwang:2008aa} [and Eq.~\eqref{phonon}] assumes that the dominant contribution to the scattering comes from the longitudinal acoustic phonons. A more recent study~\cite{Kaasbjerg:2012aa} shown, that the transverse acoustic phonons dominate at low temperatures. As a result, the resistivity obeys the power law $\rho_{ph}\propto T^\alpha$ with $\alpha\sim 6$, even in the absence of screening~\cite{Hwang2007PRB75205418}, which can additionally impair the impact of the phonon-related scattering.
The screening in the case of the hybrid system is a nontrivial issue, which requires separate consideration. We only note, that here the screening can likely be disregarded for certain $l$.
Thus we conclude, that at low temperatures, the $T$-dependence of resistivity due to bogolon-mediated scattering events is fundamentally different from the phonon case. Since the bogolons have smaller temperature exponent $T^4$ than phonons, we envisage the former to dominate (at $T\ll T_{BG}$). 

It should be emphasized that we do not have to put $\tilde{l}\ll 1$. However, the general case does not allow for analytical extraction of the temperature dependence of resistivity out of the integration, thus requiring a numerical approach.
%


\textit{Numerical treatment.---}To build the plots in full temperature range, we use Eqs.~\eqref{eq.6},~\eqref{eq.7}, and~\eqref{eq.12} and parameters, typical for GaAs-based structures: $\epsilon=12.5\epsilon_0$, where $\epsilon_0$ is a vacuum permittivity, $M=0.52m_0$, where $m_0$ is the free electron mass, $d=10$ nm, $l=10$ nm, and $v_F=10^8$~cm/s~\cite{Castro-Neto:2009aa,Das-Sarma:2011aa}. 

Figure~\ref{fig:3} shows the inverse energy--dependent relaxation time as a function of energy for different temperatures. We thus compare the bogolon-mediated scattering with the acoustic phonon-assisted relaxation under the conditions~\cite{[{The formulars are taken from \cite{Hwang:2008aa} with graphene mass density $\rho=7.6\times 10^{-8}$~g$/$cm$^2$; phonon velocity $v_{ph}=2\times 10^6$~cm/s; deformation potential $D=6.8$~eV from \cite{Kaasbjerg:2012aa}; electron density $n=10^{12}$~cm$^{-2}$ }]phonon_para}.
There are some similarities between the bogolon- and phonon-mediated processes. In both cases, the inverse lifetime grows with the increase of temperature due to the increase of the number of fermions and bosons (bogolons or phonons) in the system. We also observe low-temperature dips at the Fermi energy, which are due to the sharpening of the Fermi surface.

Nevertheless there is a conceptual difference between the two principal channels of scattering, tracing its origin to the mechanisms of electron-phonon and electron-bogolon interaction. 
The former is stemming from the crystal lattice deformation potential theory, while electron-bogolon interaction has electric nature, and the matrix element contains the Coulomb interaction term. 

Figure~\ref{fig:4} demonstrates the behavior of the graphene resistivity as a function of temperature for different condensate densities and interlayer spacings. We also compare it with the phonon-mediated resistivity.
All the curves show $\sim T^4$ dependence at low temperatures and $\sim T$ at high temperature. 
Thus the principal behavior of resistivity is deceptively similar to the phonon-assisted case, reported in~\cite{Hwang:2008aa}. 
In the case of bogolons, different $n_c$ affect the sound velocity, and the Bloch-Gr\"{u}neisen temperature changes correspondingly: $T_{BG} \approx 54$, $190$, and $540$~K for the densities $n_c=10^{10}$, $10^{11}$, and $10^{12}$~cm$^{-2}$, respectively.
That is why we have a better agreement between the numerical results and $T^4$-analytics in the high density regime.

It should be noted, however, that given the temperature required for the state-of-the-art exciton condensation, it seems impossible to reach the $\sim T$ limit, unless we choose $n_c=10^{10}$ cm$^{-2}$ or smaller to check the fundamental high-$T$ asymptotics.
Figure~\ref{fig:4}(b) also shows, that decreasing $l$, we can increase the strength of the Coulomb interaction and then the resistivity of graphene increases.

%
%
%
%
%
%


\textit{In conclusion,} we have studied the finite--temperature electron conductivity in graphene, coupled with a two-dimensional dipolar exciton gas via the Coulomb interaction. We have calculated the energy--dependent relaxation time of electrons, accompanied by the emission and absorption of a Bogoliubov excitation.
We have further calculated the resistivity of graphene in this hybrid Bose-Fermi system and showed that bogolon-mediated scattering not only gives a significant correction to the phonon-assisted relaxation but it prevails, given specific system geometry and temperatures. 
We believe the reported results can be used to design new types of graphene-based hybrid systems, encompassing high-temperature superconductivity.


We would like to thank M.~Fistul and D.~Gangardt for useful discussions. We have been supported by the Institute for Basic Science in Korea (Project No.~IBS-R024-D1) and the Russian Science Foundation (Project No. 17-12-01039).



%
%
%
%


%
%
%
%

\bibliography{library}
\bibliographystyle{apsrev4-1}

\end{document}